\documentclass[12pt,a4paper,fleqn]{article}
\usepackage{times}
\usepackage{cite}

\usepackage[]{graphicx}
\usepackage{bm}
\usepackage{amsmath}
\usepackage{epstopdf}
\usepackage{caption}
\usepackage{mathptmx} 
\begin{document}
\begin{center}
\large
\textbf{Motion of a thin elliptic plate under symmetric and asymmetric orthotropic friction forces.}
\end{center}
\begin{center}
O. A.Silantyeva \footnote{Corresponding author\quad E-mail:~\textsf{olga.silantyeva@gmail.com}}, N.N. Dmitriev
\end{center}
\begin{center}
\textit{Department of Mathematics and Mechanics, Saint-Petersburg State University\\
198504 Universitetski pr.28, Peterhof, Saint-Petersburg, Russia}
\end{center}

\begin{abstract}
  Anisotropy of friction force is proved to be an important factor in various contact problems. We study dynamical behavior of thin plates with respect to symmetric and asymmetric orthotropic friction. Terminal motion of plates with circular and elliptic contact areas is mainly analyzed. Evaluation of friction forces for both symmetric and asymmetric orthotropic cases are shown. Regular pressure distribution is considered. Differential equations are formulated and solved numerically for a number of initial conditions. Examples show significant influence of friction force asymmetry on the motion.
\end{abstract}

\normalsize
\section{Introduction}
\label{sec:intro}

Anisotropy of friction force is an important factor in contact problems. Surfaces of wide number of materials (crystals,  composited, polymers, wood) are anisotropic due to their internal properties. Engineering materials usually are manufactured in the way that oriented surface roughness textures appear. Smart materials with directional asymmetry are developed \cite{Bafekrpour2015}. Using atomic force microscope (AFM) in \cite{Campione2012} way of obtaining frictional hodographs is shown. This information is related to intensity and symmetry of friction phenomenon. Contact  process with wear, deformation, surface evolution, thermal and chemical variations, impact frictional behavior of materials. A review \cite{Zmitrowicz2005} discusses factors influencing friction forces and  approaches of  friction force modeling. The paper \cite{Zmitrowicz2010} presents examples of friction forces with respect to contact stresses calculation in various computational tasks.
 
Influence of anisotropy at contact interface has been widely investigated experimentally and theoretically during last decades by \cite{Dmitriev2002,Farkas2003,Konyukhov2008,Weidman2007} and others. Examples of centrosymmetric and non-centrosymmetric friction are presented in \cite{Zmitrowicz1992_ex}.  In \cite{Antoni2007} authors  proposed a generalized Coulomb-like friction law and set up a series of experiments with parallelepiped test speciments with asymmetric surface texture. 
 
This study deals with symmetric and asymmetric orthotropic friction. We investigate dynamical behavior of sliding and spinning disks on anisotropic surfaces. In  \cite{Weidman2007} some experimental and theoretical results regarding terminal motion of sliding spinning disks are presented. Sliding and spinning motions are  also considered in \cite{Farkas2003}. However, both papers assume isotropic friction force. We attempt to finalize our work done during last years. We investigated a  circular area with symmetric orthotropic friction and uniform pressure distribution in \cite{Dmitriev2009_4}, an elliptic area with symmetric orthotropic friction and uniform pressure distribution in \cite{Silantyeva2016_vestnik} and linear pressure distribution in \cite{Silantyeva2014}, a mass point with asymmetric friction in \cite{Dmitriev2013}, a mass point and elliptic plate with asymmetric friction in \cite{Silantyeva2016_gdansk} and a ring with asymmetric friction in \cite{Dmitriev2015}. 

In the paper  we present descriptions, regarding the developed theory for circular and elliptic thin plates  under uniform pressure distribution. Effect of asymmetry of friction force is taken into account and compared with symmetric case. Asymmetry of friction is used in omni-directional vehicles \cite{Ishigami2012} and robotics \cite{Carbone2009}.  Elliptical contact area, which appear in railway problems (see \cite{Piotrowski2005}), in multi-body dynamics during analysis of foot motion (see \cite{Lopes2015}) and other situations is considered. All equations are evolved for this contact domain as a generalized form which include circular area as a test base. Main results are presented for the elliptic contact area.

\section{Formulation of the problem}\label{sec:problem}

\subsection{Friction law}\label{subsec:friclaw}

Let us consider terminal motion  of a thin plate on a horizontal plane with anisotropic friction force. 
Anisotropic friction force ${\bf T}$ at a point $ M $ of a moving body according to \cite{Zmitrowicz1989} can be written in the following form:

\begin{equation}\label{eq:fric_law}
{\bf T} = - p_M {\mathcal{F}}(M) \frac{{\bf v}}{|{{\bf v}}|} ,\quad \mathcal{F}(M) = 
\begin{pmatrix}
 f_{x} & f \\ 
-f & f_{y} 
\end{pmatrix},
\end{equation}
here $p_M$ is a normal pressure at the point~$M$,
${\mathcal{F}}(M)  $ is a friction matrix written with respect to a stationary coordinate system $Oxy$ (see \cite{Dmitriev2002}), ${\bf v}$ is a velocity vector of the point~$M$.

Friction is {\it symmetric orthotropic} in case the  friction matrix $\mathcal{F}(M)$ is a tensor which components are constant and do not depend on the orientation of contacting areas. This approximation is possible for the case when hardness of one plane is greater than hardness of another one or one of the contact bodies has isotropic frictional properties. If hardness of each material of the contacting pair is similar we should use a more complicated law for friction force (see \cite{Dmitriev2002}).

Friction is {\it asymmetric orthotropic} if in the friction matrix components differ in negative and positive directions of sliding with $f_{x+} \ge f_{x-}$, $f_{y+} \ge f_{y-}$. Thus, in (\ref{eq:fric_law}) we have:
$$f_x = \begin{cases} f_{x+}, & v_x \ge 0\\ f_{x-}, & v_x < 0 \end{cases}\quad and  \quad f_y = \begin{cases} f_{y+}, & v_y \ge 0\\ f_{y-}, & v_y < 0 \end{cases}, $$
here $v_x, v_y$ are projections of velocity vector in $Oxy$. 

For both cases term {\it orthotropic} means an assumption that $f=0$.

\subsection{Equations of motion}\label{subsec:eq}

Let us introduce a moving coordinate system $C\xi\eta\zeta$, associated with the plate. For the elliptic plate this coordinate system is associated with the principal axes of it. Axis $C\zeta$ is perpendicular to the plane of sliding.  Stationary coordinate system $Oxy$ is selected thus, that the friction matrix has the form (\ref{eq:fric_law}) and axes $Ox$ and $Oy$ are in the sliding plane. Let $\varphi$ be an angle between $Ox$ and $C\xi$, $\vartheta$ is an  angle between axis $Ox$ and ${\bf v_C}$ which is a velocity of a center of mass of the plate.
\begin{equation}
{\bf v_C} = v_C (\cos\vartheta {\bf i} + \sin\vartheta {\bf j}), 
\end{equation}
where $v_C$ is  a velocity value, ${\bf i}, \quad {\bf j}$ are unit vectors of axes $Ox, \quad Oy$. Vector of an angular velocity is ${\bm{\omega}} = \omega {\bf k}$, where $\omega = \dot{\varphi}$, ${\bf k}$ is a unit vector of axis $Oz$.

Euler equation  ${\bf v_M} = {\bf v_C} + {\bm{\omega}}\times{\bf CM}$ allows us to write the following statements:
\begin{equation}\label{eq:velocity}
\begin{array}{l l l}
\displaystyle
v_x = v_C \cos \vartheta - \omega y',\quad v_y = v_C\sin\vartheta + \omega x',\\ 
\displaystyle
x' = \xi\cos\varphi - \eta \sin\varphi,\quad y' = \xi\sin\varphi+\eta\cos\varphi,\\
\displaystyle
h = \eta\cos(\vartheta-\varphi) - \xi\sin(\vartheta-\varphi),\\
\displaystyle
v_M = \sqrt{v_C^2 + \omega^2(\xi^2 + \eta^2) - 2v_C\omega h}.
\end{array}
\end{equation}

With the anisotropic friction law (\ref{eq:fric_law}) and equations (\ref{eq:velocity}) we can write projections of the total friction force vector ${\bf T}$ and the total friction moment ${\bf M}$ in the form:

\begin{equation}\label{eq:forces_xy}
\begin{array}{l l l}
\displaystyle
T_x = \iint\limits_{\Omega} \tau_x d\xi d\eta, \quad T_y = \iint \limits_{\Omega} \tau_y d\xi d\eta,\\
\displaystyle 
M_{C\zeta} = \iint\limits_{\Omega}(\tau_y x' - \tau_x y')d\xi d\eta,\\
\displaystyle
\tau_x = -f_x p(\xi,\eta)\frac{v_x(\xi,\eta)}{v_M(\xi,\eta)},\quad \tau_y = -f_y p(\xi,\eta)\frac{v_y(\xi,\eta)}{v_M(\xi,\eta)},\\ 

\end{array}
\end{equation}
where $\Omega$ is an integration area.

Equations of motion in the Frenet-Serret frame with respect to (\ref{eq:forces_xy}) are as follows:
\begin{equation}\label{eq:frenet} 
\begin{array}{l l l}
\displaystyle
m\dot v_C = T_{\tau} = T_x\cos\vartheta + T_y\sin\vartheta,\\ 
\displaystyle
m v_C \dot\vartheta = T_n = -T_x\sin\vartheta + T_y\cos\vartheta,\\ 
\displaystyle
I\dot\omega = M_{C\zeta},
\end{array}
\end{equation}
where $m$ is a mass of the plate, $I$ is a plate's inertia moment about  $C\zeta$, $T_{\tau}$ and $T_n$ are projections of a friction force vector ${\bf T}$ on tangential and normal axes respectively, $M_{C\zeta}$ is a friction moment about axis $C\zeta$.

Let's rewrite the system  (\ref{eq:frenet}) in the dimensionless form using following relations:
$$I = ma^2 I^*, \quad \xi = a\xi^*, \quad \eta = a\eta^*, \quad v_C = v^*_C\sqrt{ag},$$
$$\omega = \omega^*\sqrt{\frac{g}{a}}, \quad t = t^*\sqrt{\frac{a}{g}}, \quad  \dot\vartheta = \frac{d\vartheta}{dt^*}\sqrt{\frac{g}{a}}, \quad p = p^*\frac{mg}{S} $$
and let's introduce a variable  $\displaystyle \beta = \frac{v_C}{\omega} = a\beta^*$ and a parameter $\mu = f_y - f_x$.
In these equations parameter $a$ is measurable, it is the length of the largest line from point $C$ to the area's boundary, $S$ is a contact area.

Equations (\ref{eq:frenet}) in dimensionless form (asterisks are omitted):

\begin{equation}\label{eq:move_frenet_dimless}
\begin{array}{l l l l l l l}
\displaystyle
\frac{dv_C}{dt} = -\iint\limits_{\Omega} p(\xi,\eta) \left[\frac{\beta(f_x + \mu\sin^2\vartheta) + f_x s_1+\mu s_3 + f s_2}{s}\right] d\xi d\eta,\\ 
\\
\displaystyle
v_C\frac{d\vartheta}{dt} = -\iint\limits_{\Omega} p(\xi,\eta) \left[\frac{\beta (\mu \sin\vartheta\cos\vartheta-f) + f_x s_2 +\mu s_4 - f s_1}{s}\right] d\xi d\eta,\\ 
 \\
 \displaystyle
\frac{d\omega}{dt} = -\iint\limits_{\Omega}\frac{p(\xi,\eta)}{I}  \left[\frac{\beta (f_x s_1  + \mu s_3 - f s_2)+f_x(\xi^2+\eta^2) + \mu s_0^2}{s}\right] d\xi d\eta,
\end{array}
\end{equation}
where
\begin{equation}
\begin{array}{l l l l l l l}\nonumber
\displaystyle
 s = \sqrt{\beta^2 + \xi^2+\eta^2  + 2\beta s_1 },\quad 
 s_0 = \xi\cos\varphi - \eta\sin\varphi,\\ \nonumber
 s_1 = \xi\sin(\vartheta-\varphi)-\eta\cos(\vartheta-\varphi),\quad 
 \displaystyle 
 s_2 = \xi\cos(\vartheta-\varphi)+\eta\sin(\vartheta-\varphi),\\ \nonumber 
s_3 = \xi\cos\varphi\sin\vartheta - \eta\sin\varphi\sin\vartheta,\quad 
\displaystyle s_4 = \xi\cos\varphi\cos\vartheta - \eta\sin\varphi\cos\vartheta.
 \end{array}
\end{equation} 

System of equations (\ref{eq:move_frenet_dimless}) is  general. It is possible to numerically evaluate this system directly. However, in most cases it is better to integrate forces in  the right part of the system at least once -- it accelerates calculations and simplifies analysis. We will study later uniform pressure distribution and only orthotropic case.

\section{Friction force evaluation}\label{sec:force}
\subsection{Symmetric orthotropic friction}\label{subsec:sym_fric}

\begin{figure}[!t]
\begin{center}
\includegraphics[width=9cm]{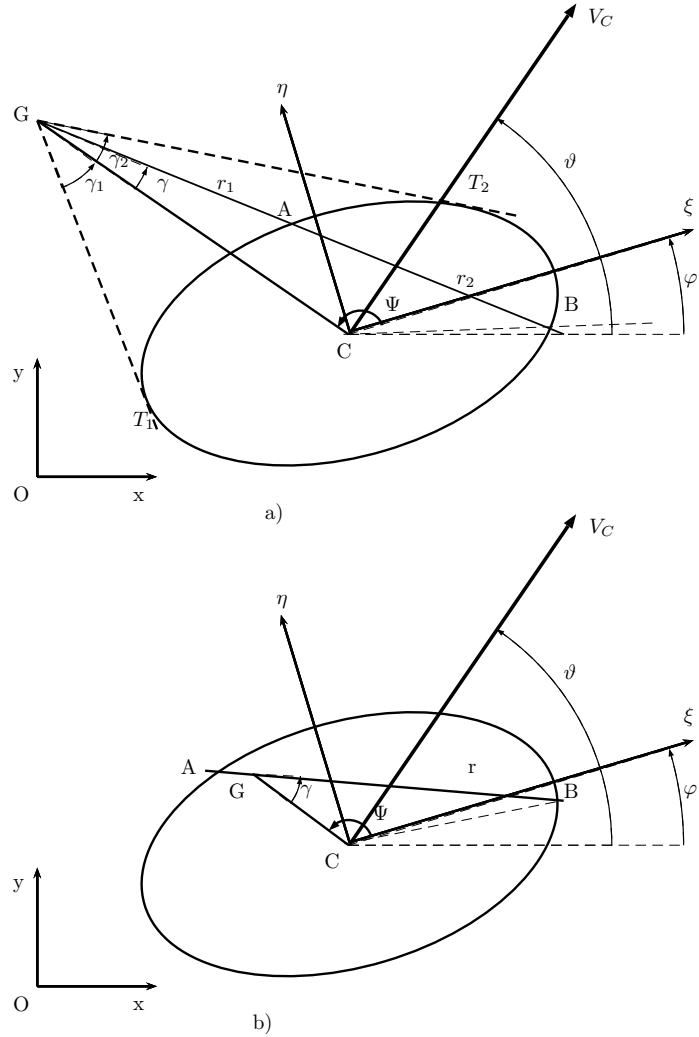}
   \caption{Coordinate system. Method A.I.~Lurye: instantaneous velocity center, a) outside area, b) inside area}
   \label{fig:1:luryegeom}
\end{center}
\end{figure}

We will evaluate friction forces using method developed by A.I.~Lurye in \cite{Lurye2002}. Let us introduce polar coordinate system, which origin is in the instantaneous velocity center $G$, polar axis is a ray from simultaneous velocity center through plate center $C$, $\gamma$ is a polar angle. We will differ two cases of simultaneous velocity center position: inside and outside area, covered by plate, see figure~\ref{fig:1:luryegeom}.

Velocity vector is the following:
\begin{equation}\nonumber
{\bf v} = v (\cos(\vartheta+\gamma){\bf i} +  \sin(\vartheta+\gamma){\bf j}),
\end{equation}
where $v$ is a velocity value, ${\bf i}, {\bf j}$ are  unit vectors of coordinate system axes.

Vector of elementary friction force is: 
\begin{equation}\label{eq:elemfricforce}
\bm{\tau} = -p(f_x \cos(\vartheta+\gamma){\bf i} + f_y \sin(\vartheta+\gamma){\bf j}). 
\end{equation}

Friction force and moment taking into account (\ref{eq:elemfricforce}) in stationary coordinate system are the following:

\begin{equation} \label{eq:force_lurye_main}
\begin{array}{l l l}
\displaystyle
T_x = -p f_x \int \limits_{\gamma_1}^{\gamma_2}\int \limits_{r_1(\gamma)}^{r_2(\gamma)}  \cos(\vartheta+\gamma) r dr d\gamma,\\
\displaystyle
T_y = -p f_y \int \limits_{\gamma_1}^{\gamma_2}\int \limits_{r_1(\gamma)}^{r_2(\gamma)} \sin(\vartheta+\gamma)  r dr d\gamma,\\
\displaystyle
M_G = -p \int \limits_{\gamma_1}^{\gamma_2} \int \limits_{r_1(\gamma)}^{r_2(\gamma)} (f_x + \frac{\mu}{2} -\frac{\mu}{2}\cos(2\vartheta +2\gamma))  r^2 dr d\gamma,\\
\displaystyle
M_C = M_G - CG \cdot  p \int \limits_{\gamma_1}^{\gamma_2} \int \limits_{r_1(\gamma)}^{r_2(\gamma)} (\frac{\mu}{2}\cos(\gamma) - \frac{\mu}{2}\cos(2\vartheta + \gamma)+f_x\cos(\gamma))  r dr d\gamma.\\
\end{array}
\end{equation}

We assume that the plate has elliptical shape with semi-axes $a$ and
$b$, where $a$ is a major semi-axis, than
$$m = \rho \pi a b, \quad \kappa =
\sqrt{1-e^2},\quad I =
\frac{\rho\pi\kappa a^4(1+\kappa^2)}{4},\quad p =
\frac{mg}{\pi ab},$$  where $m$ is a mass of the plate, $\rho$ is a mass density of the plate, $e$ is an ellipse eccentricity, $p$ is a value of uniformly distributed pressure, $g$ is the acceleration of free fall.

Let's find integration ranges in (\ref{eq:force_lurye_main}). We introduce angle $\displaystyle \Psi = \frac{\pi}{2} + \vartheta-\varphi$ (see~fig.~\ref{fig:1:luryegeom}). If the point  $G$ is inside elliptical plate, distance $r$ from $G$ to the border of contacting area $B$ can be found from:

\begin{equation}\label{eq:gb}\nonumber
\displaystyle
\frac{(GB\cos(\pi - (\Psi+\gamma)) - \xi_G)^2}{a^2} + \frac{(GB\sin(\pi - (\Psi+\gamma)) - \eta_G)^2}{b^2} = 1,
\end{equation}
here coordinates of instantaneous velocity center are: $$\xi_G = x_G \cos\varphi + y_G \sin\varphi, \quad \eta_G = -x_G \sin\varphi + y_G\cos\varphi,$$ $$\displaystyle x_G = -\frac{v}{\omega}\sin\vartheta, \quad y_G = \frac{v}{\omega}\cos\vartheta.$$

Thus, we receive:
\begin{equation}\label{eq:r_in}
\displaystyle
r = GB = \frac{\lambda_1 + ab D_1}{\lambda_2},
\end{equation}
where $$\lambda_1 = b^2\xi_G\cos(\Psi+\gamma) + a^2\eta_G\sin(\Psi+\gamma),$$ $$\lambda_2 = b^2\cos^2(\Psi+\gamma) + a^2\sin^2(\Psi+\gamma),$$ $$D_1  = \sqrt{(b^2-\eta_G^2)\cos^2(\Psi+\gamma) + (a^2-\xi_G^2)\sin^2(\Psi+\gamma) + \xi_G\eta_G\sin(2(\Psi+\gamma))}.$$
Angle $\gamma$ in equation (\ref{eq:r_in}) takes values from  $0$ to  $2\pi$ (see~\cite{Rozenblat2006}).

The same calculations for the point $G$ outside the area lead to the following:
\begin{equation} \label{eq:r_out}
\begin{array}{l l}
\displaystyle
r_1 = GA = \frac{\lambda_1 - ab D_1}{\lambda_2},\\
\displaystyle
r_2 = GB = \frac{\lambda_1 + ab D_1}{\lambda_2}.\\
\end{array}
\end{equation}

Let's find angles $\gamma_1$ and $\gamma_2$ for that case.
Points where line $GA$ intersects ellipse in coordinate system $C\xi\eta$ can be found from the system of equations:
\begin{equation} \label{eq:range_r}\nonumber
\begin{array}{l l}
\displaystyle 
\frac{\xi^2_A}{a^2} + \frac{\eta^2_A}{b^2} = 1,\\
\eta_A = \xi_A k + \sigma,\\
\end{array}
\end{equation} 
where $$\sigma = \eta_G - \xi_G k, \quad k = \tan(\frac{\pi}{2} + \vartheta -\varphi +\gamma).$$ 
The line will be tangent to ellipse if the following equation is satisfied:
$$\sigma^2 = a^2k^2+b^2.$$ 
We receive equation for parameter $k$:
$$(\xi^2_G - a^2) k^2 - 2\eta_G\xi_G k + \eta^2_G - b^2 = 0.$$ 
Thus:
$$k_{1,2} = \frac{\eta_G\xi_G \pm \sqrt{\xi^2_Gb^2 + a^2\eta^2_G - a^2b^2}}{\xi^2_G - a^2},$$ 
and, finally, 
\begin{equation}\label{eq:gamma_range}
\gamma_{1,2} = \arctan(k_{1,2})  - \frac{\pi}{2} - \vartheta + \varphi.
\end{equation}

We can rewrite forces in dimensionless form taking into account the following relations:
\begin{equation}\label{eq:dimless} \nonumber
\begin{array}{l l l l}
\displaystyle
\quad I =
\rho\pi\kappa a^4 I^*,\quad \xi =
a\xi^*, \quad \eta = a \kappa\eta^*, \quad 
\beta = \beta^* a, \quad p = p^* \frac{mg}{\pi a b},
\end{array}
\end{equation} 
here asterisks are dedicated to dimensionless variables: $I^*$ is a dimensionless inertia moment, $\xi^*, \eta^*$ are dimensionless coordinates (asterisks for these variables are omitted later).

Thus, we can write components of total friction force and total moment in the case, when instantaneous velocity center lies inside the contact area:
\begin{equation} \label{eq:sym_force_in}
\begin{array}{l l l}
\displaystyle
T^*_x = - f_x \int \limits_{0}^{2\pi} \cos(\vartheta+\gamma)\left(\frac{\kappa\left(\lambda_1^*+ D_1^*\right)^2}{2 \lambda_2^{*2}}\right)d\gamma,\\

\displaystyle
T^*_y = - f_y \int \limits_{0}^{2\pi} \sin(\vartheta+\gamma)\left(\frac{\kappa\left(\lambda_1^*+ D_1^*\right)^2}{2 \lambda_2^{*2}}\right)d\gamma,\\ 

\displaystyle
M^*_G = - \int \limits_{0}^{2\pi} (f_x + \frac{\mu}{2} -\frac{\mu}{2}\cos(2\vartheta +2\gamma)) \left(\frac{\kappa^2\left(\lambda_1^* + D_1^*\right)^3}{3\lambda_2^{*3}}\right) d\gamma,\\

\displaystyle
M^*_C = M_G^* - \int \limits_{0}^{2\pi} \beta(\frac{\mu}{2}\cos(\gamma)-\frac{\mu}{2}\cos(2\vartheta+\gamma)+ f_x\cos(\gamma)) \left(\frac{\kappa\left(\lambda_1^* + D_1^*\right)^2}{2\lambda_2^{*2}}\right) d\gamma.\\
\end{array}
\end{equation} 
And in the case, when instantaneous velocity center is outside the contact area:

\begin{equation} \label{eq:sym_force_out}
\begin{array}{l l l}
\displaystyle
T^*_x = - f_x \int \limits_{\gamma_1}^{\gamma_2} \cos(\vartheta+\gamma)\left( \frac{2\kappa D_1^*\lambda_1^*}{\lambda_2^{*2}}\right) d\gamma,\\
\displaystyle
T^*_y = - f_y \int \limits_{\gamma_1}^{\gamma_2} \sin(\vartheta+\gamma)\left( \frac{2\kappa D_1^*\lambda_1^*}{\lambda_2^{*2}}\right)  d\gamma ,\\
\displaystyle
M^*_G = - \int \limits_{\gamma_1}^{\gamma_2} (f_x + \frac{\mu}{2} -\frac{\mu}{2}\cos(2\vartheta +2\gamma)) \left(\frac{\kappa^2\left(2D_1^{*3} + 6D_1^*\lambda_1^{*2}\right)}{3 \lambda_2^{*3}}\right) d\gamma,\\
\displaystyle
M^*_C = M_G^* -  \int \limits_{\gamma_1}^{\gamma_2} \beta (\frac{\mu}{2}\cos(\gamma)-\frac{\mu}{2}\cos(2\vartheta+\gamma)+ f_x\cos(\gamma)) \left( \frac{2\kappa D_1^*\lambda_1^*}{\lambda_2^{*2}}\right) d\gamma,\\
\end{array}
\end{equation} 
where
$$\lambda_1^* = \kappa \xi_G \cos(\Psi+\gamma) + \eta_G\sin(\Psi+\gamma),$$
$$\lambda_2^* = \kappa^2\cos^2(\Psi+\gamma) + \sin^2(\Psi+\gamma),$$
$$D_1^*  = \sqrt{\kappa^2(1-\eta_G^2)\cos^2(\Psi+\gamma) + (1-\xi_G^2)\sin^2(\Psi+\gamma) + \kappa\xi_G\eta_G\sin(2(\Psi+\gamma))}.$$

\subsection{Asymmetric orthotropic friction}\label{subsec:asym_fric}

Equations (\ref{eq:sym_force_in}) and (\ref{eq:sym_force_out}) achieved in the previous section for symmetric orthotropic friction are suitable for asymmetric case as well. However, we should evaluate another integration ranges. In this case it is important to know directions of sliding velocities. Each area on figure \ref{fig:splitgeom} corresponds to different  cases of velocities orientation and, thus, different friction coefficients.

We have to find rules for positioning instantaneous velocity center $G$  in each area.  Let's introduce points $p1, p2, p3, p4$, which are coordinates of  tangents to ellipse parallel to axes $Cx$ and $Cy$.

In system $C\xi\eta$:
equation of line $p_3$ is $\eta = k_3 \xi + l_3$, with $k_3 = -\tan\varphi$. Coordinates of contact point in system  $C\xi\eta$ can be found from the following relations:
\begin{equation}\nonumber
\begin{array}{lll}
\displaystyle
\eta = -\tan\varphi \xi + l_3,\\
\displaystyle
\frac{\xi^2}{a^2} + \frac{\eta^2}{b^2} = 1.
\end{array}
\end{equation}
Taking $l_3 = \sqrt{a^2k^2_3 + b^2}$ we obtain
$$\xi_{p3} = -\frac{a^2k_3}{\sqrt{a^2k_3 + b^2}},$$ 
$$\eta_{p_3} = k_3\xi_{p3} + l_3,$$
and in system $Cxy$ we get:
\begin{equation}\nonumber
\begin{array}{lll}
\displaystyle
x_{p3} = \xi_{p3}\cos\varphi - \eta_{p3}\sin\varphi,\\
\displaystyle
y_{p3} = \xi_{p3}\sin\varphi + \eta_{p3}\cos\varphi.
\end{array}
\end{equation}
The same idea is useful for other 3 points with:
$k_1 = \cot\varphi, \quad l_1 = \sqrt{a^2k_1^2 + b^2}$, $k_2 = \cot\varphi, \quad l_2 = -l_1$, $k_4 = -\tan\varphi, \quad l_4 = -l_3$.

Now take a look at area 7 (see table \ref{tab:asym_areas_p2}). It is divided into 4 parts. Each part contains points with velocities directed to the same quadrant. That means that for each of the part coefficients of friction remain constant. So projections of the friction force and the moment are:
\begin{equation} \label{eq:asym_force_in}
\begin{array}{l l l}
\displaystyle
T^*_x = - \sum \limits_{i=0, j=1}^{i=4, j=5}\int \limits_{\psi_i}^{\psi_j} f_x^{ij}\cos(\vartheta+\gamma)\left(\frac{\kappa\left(\lambda_1^*+ D_1^*\right)^2}{2 \lambda_2^{*2}}\right)d\gamma,\\

\displaystyle
T^*_y = - \sum \limits_{i=0, j=1}^{i=4, j=5} \int \limits_{\psi_i}^{\psi_j} f_y^{ij} \sin(\vartheta+\gamma)\left(\frac{\kappa\left(\lambda_1^*+ D_1^*\right)^2}{2 \lambda_2^{*2}}\right)d\gamma,\\

\displaystyle
M^*_G = - \sum \limits_{i=0, j=1}^{i=4, j=5}\int \limits_{\psi_i}^{\psi_j} (f_x^{ij} + \frac{\mu^{ij}}{2} -\frac{\mu^{ij}}{2}\cos(2\vartheta +2\gamma)) \left(\frac{\kappa^2\left(\lambda_1^* + D_1^*\right)^3}{3\lambda_2^{*3}}\right) d\gamma,\\

\displaystyle
M^*_C = M_G^* - \sum \limits_{i=0, j=1}^{i=4, j=5} \int \limits_{\psi_i}^{\psi_j} \beta(\frac{\mu^{ij}}{2}\cos(\gamma)-\frac{\mu^{ij}}{2}\cos(2\vartheta+\gamma)+ f_x^{ij}\cos(\gamma)) \left(\frac{\kappa\left(\lambda_1^* + D_1^*\right)^2}{2\lambda_2^{*2}}\right) d\gamma.\\
\end{array}
\end{equation} 

In (\ref{eq:asym_force_in}) the area integral from (\ref{eq:sym_force_in}) is partitioned to 5 integrals with constant values of friction coefficients with $\psi_0 = 0, \psi_1 = \alpha_1, \psi_2 = \alpha_2, \psi_3 = \alpha_3, \psi_4 = \alpha_4, \psi_5 = \alpha_5$ and $\sum \psi_i = 2\pi$ (see table \ref{tab:asym_areas_p2}).

Tables \ref{tab:asym_areas_p1} and \ref{tab:asym_areas_p2} show the way how equations (\ref{eq:sym_force_out}) are divided into several parts. Thus, finally, we achieve equations for friction force projections in the case of instantaneous velocity center lies outside area covered by the plate in the following form:
 
\begin{equation} \label{eq:asym_force_out}
\begin{array}{l l l}
\displaystyle
T^*_x = - \sum \limits_{i=0, j=1}^{i=z-1, j=z} \int \limits_{\psi_i}^{\psi_j}f_x^{ij}\cos(\vartheta+\gamma)\left( \frac{2\kappa D_1^*\lambda_1^*}{\lambda_2^{*2}}\right) d\gamma,\\
\displaystyle
T^*_y = - \sum \limits_{i=0, j=1}^{i=z-1, j=z}\int \limits_{\psi_i}^{\psi_j} f_y^{ij} \sin(\vartheta+\gamma)\left( \frac{2\kappa D_1^*\lambda_1^*}{\lambda_2^{*2}}\right)  d\gamma ,\\
\displaystyle
M^*_G = - \sum \limits_{i=0, j=1}^{i=z-1, j=z} \int \limits_{\psi_i}^{\psi_j} (f_x^{ij} + \frac{\mu^{ij}}{2} -\frac{\mu^{ij}}{2}\cos(2\vartheta +2\gamma)) \left(\frac{\kappa^2\left(2D_1^{*3} + 6D_1^*\lambda_1^{*2}\right)}{3 \lambda_2^{*3}}\right) d\gamma,\\
\displaystyle
M^*_C = M_G^* -  \sum \limits_{i=0, j=1}^{i=z-1, j=z}\int \limits_{\psi_i}^{\psi_j} \beta (\frac{\mu^{i,j}}{2}\cos(\gamma)-\frac{\mu^{ij}}{2}\cos(2\vartheta+\gamma)+ f_x\cos(\gamma)) \left( \frac{2\kappa D_1^*\lambda_1^*}{\lambda_2^{*2}}\right) d\gamma .\\
\end{array}
\end{equation} 

\begin{table}[!h] 
\centering
\caption{Velocities distribution and integration ranges evaluation. Part 1.}\label{tab:asym_areas_p1}
\begin{tabular}{|p{7.5cm}|p{8.4cm}|}
\hline
\begin{minipage}{.5\textwidth}
     \includegraphics[width=\linewidth]{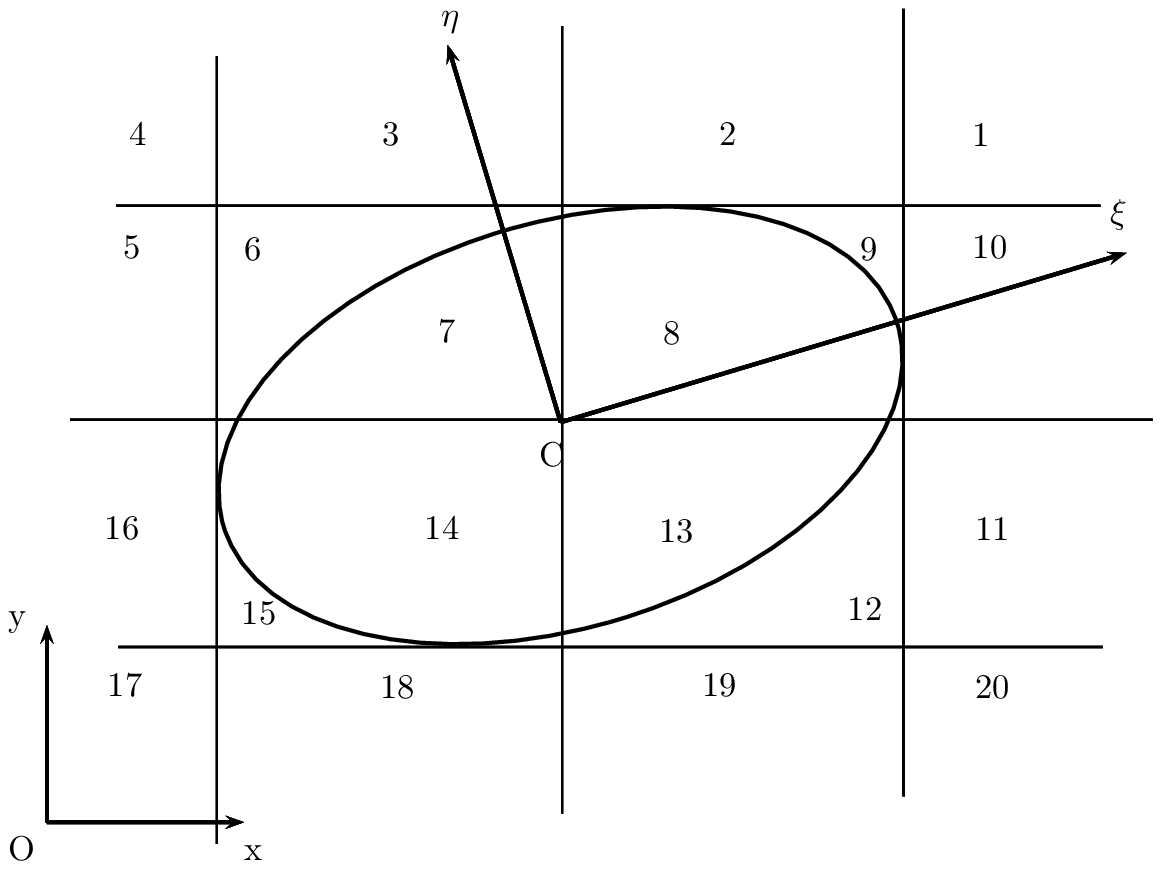}
   \captionof{figure}{Splitting geometry}
   \label{fig:splitgeom}
\end{minipage}&
Here each partition correspond to different velocity directions distribution.\\
\hline
\begin{minipage}{.5\textwidth}
     \includegraphics[width=\linewidth]{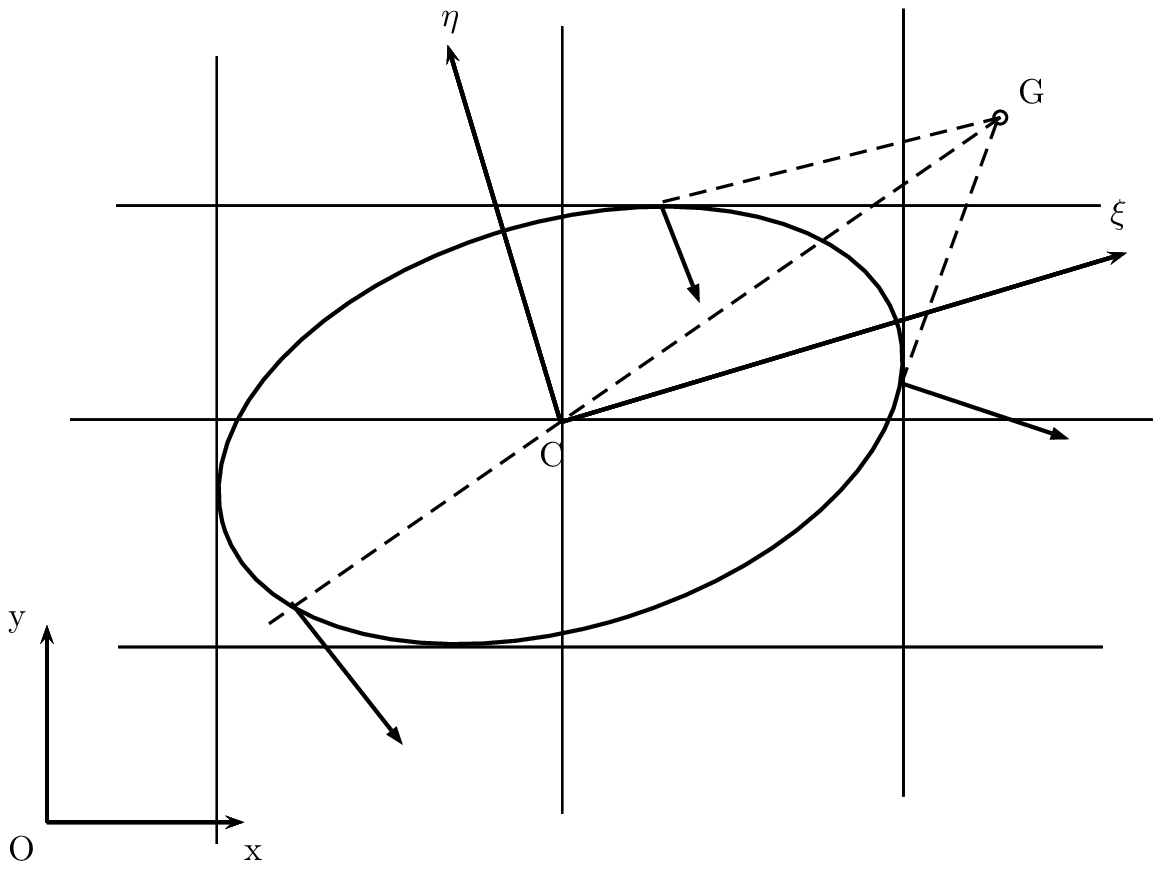}
\end{minipage}& \begin{minipage}{0.6\textwidth} Area 1, velocities oriented to the III quadrant.\\ $x_G > x_{p1}, \quad y_G > y_{p3}$ \\  $z = 1, \psi_0 = -\gamma_1, \psi_1 = \gamma_2$\\ $f_x^{01} = f_{x+}, f_y^{01} = f_{y-}$ \\ For areas 1, 4, 17, 20 all velocities are directed\\ to the same quadrant.\end{minipage}  \\
\hline
\begin{minipage}{.5\textwidth}
     \includegraphics[width=\linewidth]{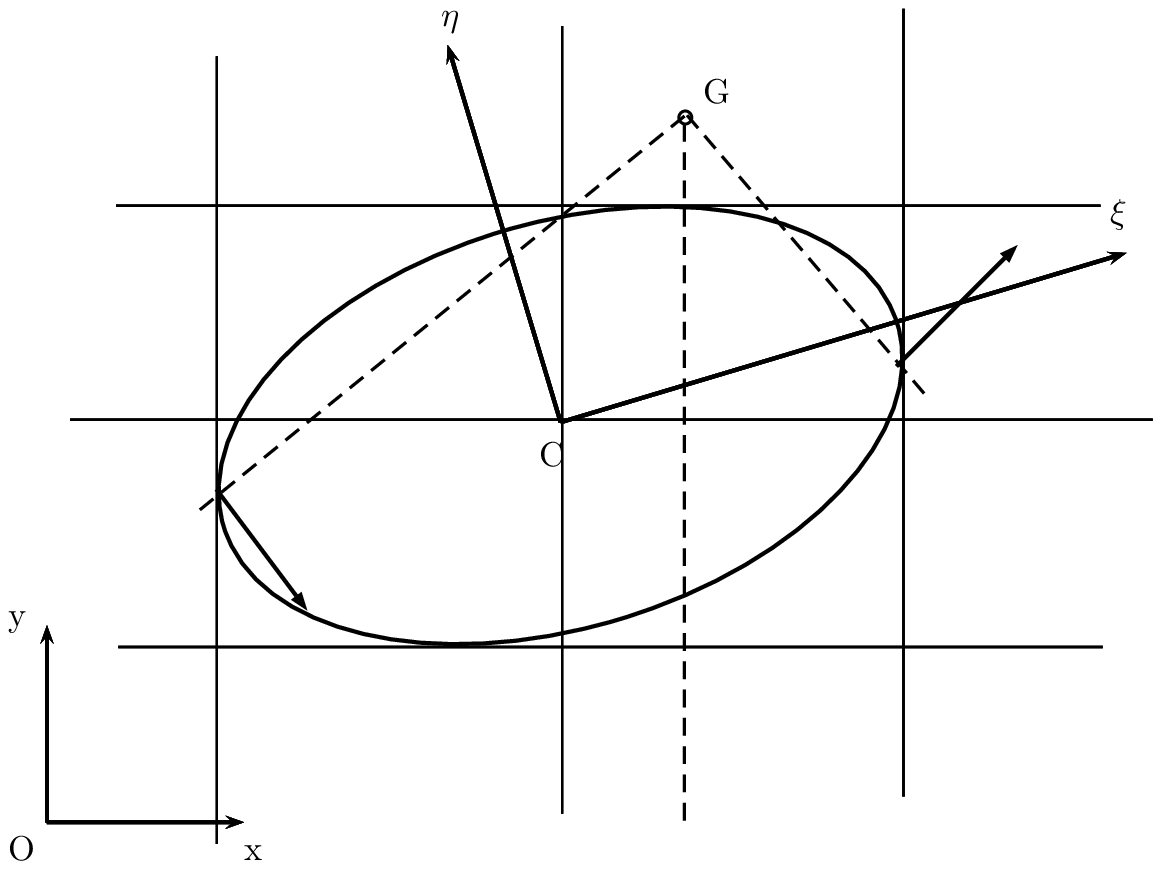}
\end{minipage} & \begin{minipage}{0.6\textwidth} Area 2, velocities oriented to the I and IV quadrants.\\ $x_G >0, \quad x_G \le x_{p1}$,\\ $ y_G > y_{p3}$, \\ $\displaystyle \alpha = \arctan\frac{x_G}{y_G}$ \\ $z=2, \psi_0 = -\gamma_1, \psi_1 = \alpha, \psi_2 = \gamma_2$\\ $f_x^{01} = f_{x+}, f_y^{01} = f_{y-},$\\ $ f_x^{12} = f_{x+}, f_y^{12} = f_{y+}$\\ For areas 2, 3, 18, 19 all velocities are directed\\ to 2 quadrants and the line separating the parts\\ is parallel to axis $Oy$.\end{minipage}\\
\hline
\end{tabular}
\end{table}
\begin{table}[!h] 
\centering
\caption{Sliding velocities distributions and integration ranges evaluations. Part 2.}\label{tab:asym_areas_p2}
\begin{tabular}{|p{7.5cm}|p{8.4cm}|}
\hline
\begin{minipage}{.5\textwidth}
     \includegraphics[width=\linewidth]{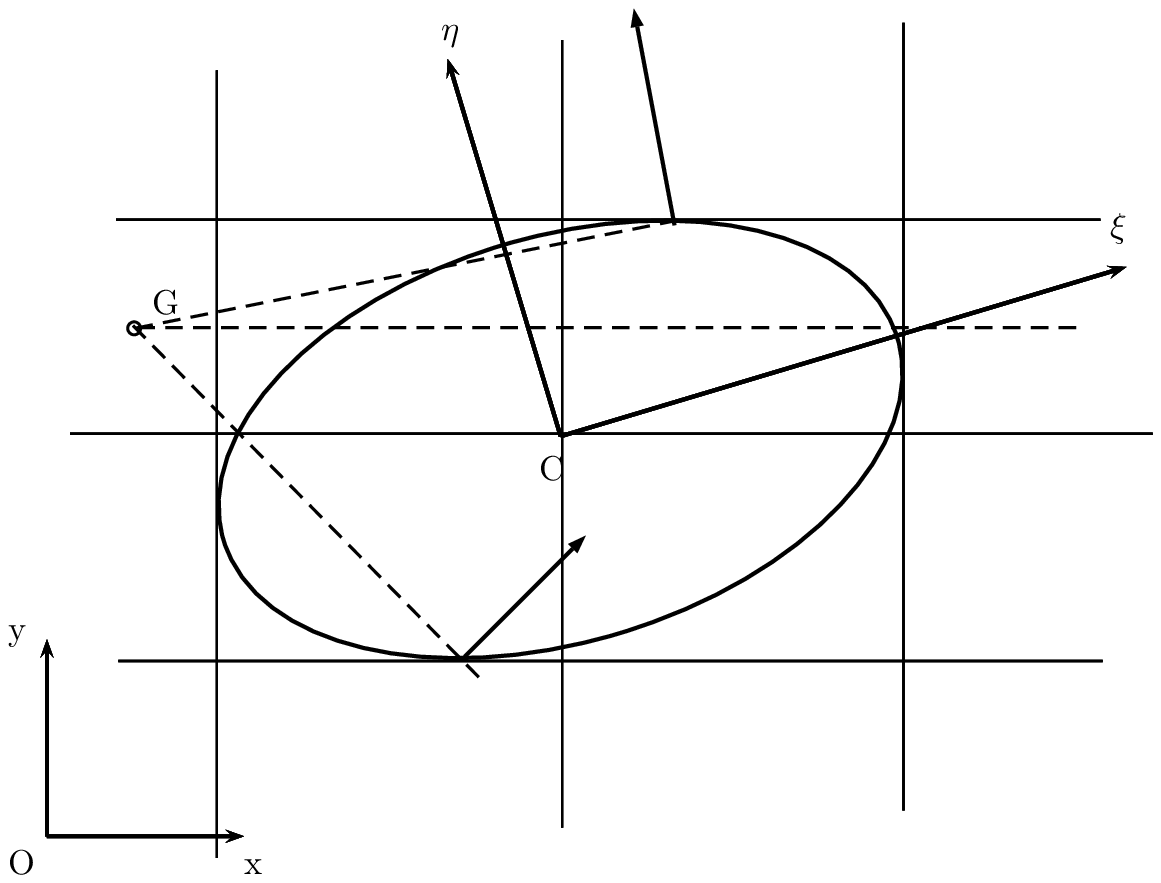}
\end{minipage}& \begin{minipage}{0.6\textwidth} Area 5, velocities oriented to the I and II quadrants.\\ $x_G \le x_{p2}$, $ y_G > 0, \quad y_G \le  y_{p3}$, $\displaystyle \alpha = \arctan\frac{y_G}{x_G}$ \\ $z = 2, \psi_0 = -\gamma_1, \psi_1 = \alpha, \psi_2 = \gamma_2$\\ $f_x^{01} = f_{x+}, f_y^{01} = f_{y+}$ \\ $f_x^{12} = f_{x-}, f_y^{12} = f_{y+}$\\ For areas 5, 10, 11, 16 all velocities are directed\\ to 2 quadrants and the line separating the parts\\ is parallel to axis $Ox$.\end{minipage} \\
\hline
\begin{minipage}{.5\textwidth}
     \includegraphics[width=\linewidth]{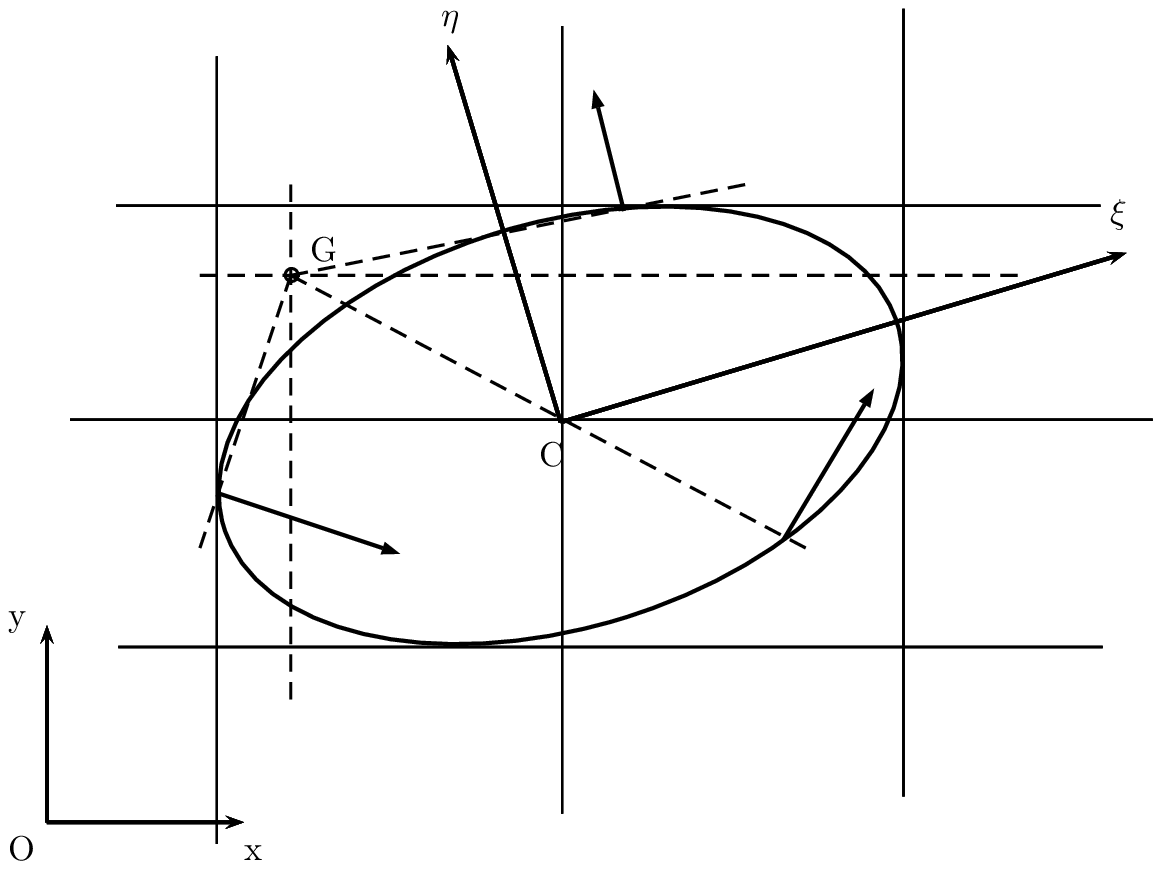}
\end{minipage}
& \begin{minipage}{0.6\textwidth} Area 6, velocities oriented to the I, II and IV quadrants. \\ $x_G \le 0, \quad x_G > x_{p2}$, $y_G >  y_{p2},\quad  y_G \le y_{p3}$, \\$\displaystyle \frac{\xi_G^2}{a^2} + \frac{\eta_G^2}{b^2} > 1$, $\displaystyle \alpha_1 = \arctan\frac{x_G}{y_G}$, $\displaystyle \alpha_2 = \arctan\frac{y_G}{x_G}$ \\ $f_x^{01} = f_{x+}, f_y^{01} = f_{y+}$ \\ $f_x^{12} = f_{x-}, f_y^{12} = f_{y+}$ \\ $f_x^{23} = f_{x+}, f_y^{23} = f_{y-}$ \\For areas 6, 9, 12, 15 all velocities are directed\\ to 3 quadrants.\end{minipage} \\
\hline
\begin{minipage}{.5\textwidth}
     \includegraphics[width=\linewidth]{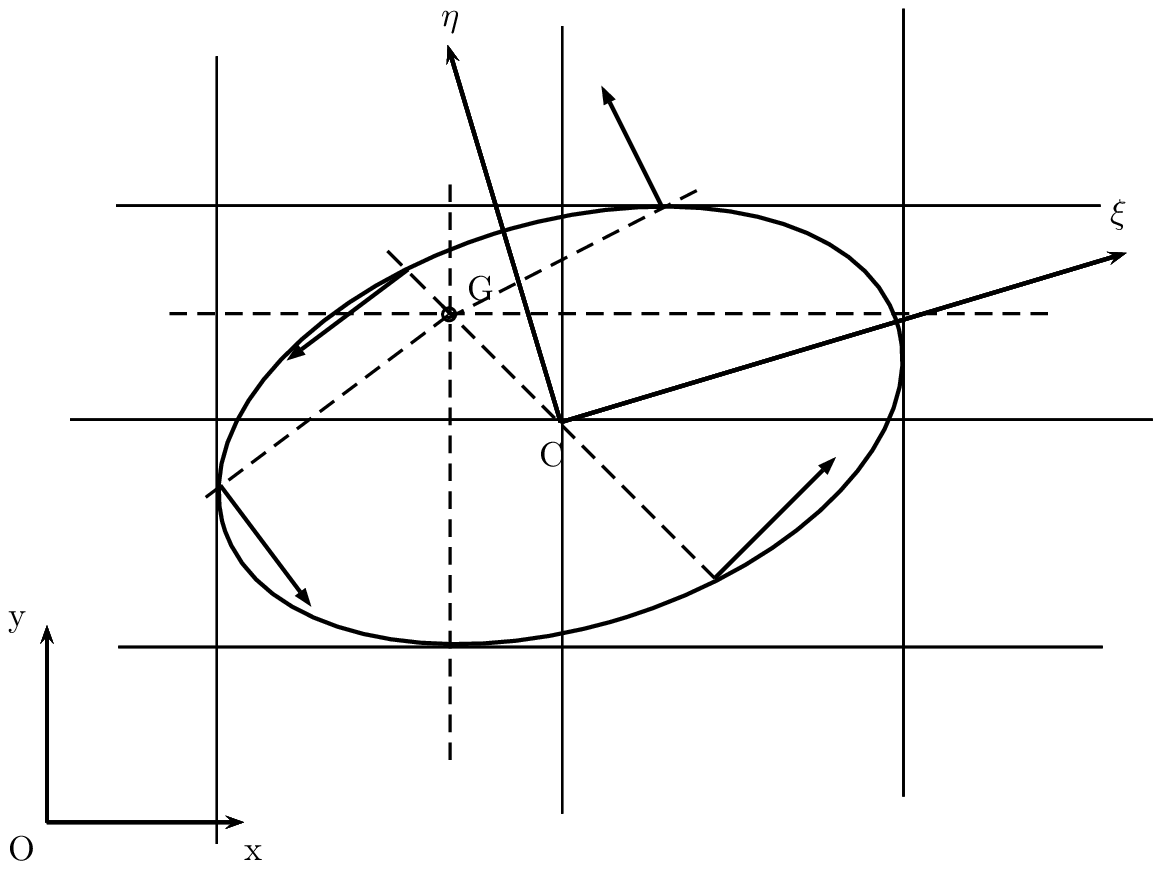}
\end{minipage}& \begin{minipage}{.6\textwidth} Area 7, velocities are directed to I, II, III, IV quadrants.\\ 
$\displaystyle \alpha_1 = \arctan\frac{y_G}{x_G}$, $\alpha_2 = \alpha_1 + \frac{\pi}{2}$,
$\alpha_3 = \alpha_2 + \frac{\pi}{2}$,\\
$\alpha_4 = \alpha_3 + \frac{\pi}{2}$,
$\displaystyle \alpha_5 = \alpha_4 + \arctan\frac{x_G}{y_G}$,\\
$x_G \le 0$, $y_G > 0$, $\displaystyle \frac{\xi_G^2}{a^2} + \frac{\eta_G^2}{b^2} \le 1$, \\
$f_x^{01} = f_{x+}, f_y^{01} = f_{y+}$ \\ $f_x^{12} = f_{x-}, f_y^{12} = f_{y+}$\\$f_x^{23} = f_{x-}, f_y^{23} = f_{y-}$\\$f_x^{34} = f_{x+}, f_y^{34} = f_{y-}$\\
For areas 6, 8, 13, 14 velocities are directed\\ to all 4 quadrants. \end{minipage}\\
\hline
\end{tabular}
\end{table}

\section{Selected results}\label{sec:res}
\subsection{General considerations}\label{subsec:res_general}

Let's divide the first equation of the system (\ref{eq:move_frenet_dimless}) by the third and let's introduce derived right part $\Phi_1(\beta,\vartheta)$:
\begin{equation}\label{eq:Phi}
\begin{array}{l l}
\displaystyle
\frac{dv_C}{d\omega} = \Phi_1(\beta,\vartheta),\\
\displaystyle
v_C\frac{d\vartheta}{dt} = T_n(\beta,\vartheta).
\end{array}
\end{equation}

The dissipative friction force has negative power, thus, motion with non-zero initial conditions terminates (the plate moves a finite period of time).  Thus, second equation in the system (\ref{eq:Phi}) allows us to write a relation:
\begin{equation}\label{eq:normal_force_rule}
T_n(\beta,\vartheta)\underset{\substack{t \to t_*\\ \vartheta\to\vartheta_*\\ \beta\to\beta_*}}\longrightarrow  0,\\ 
\end{equation}
where $\vartheta_*$ and $\beta_*$ are limit values of corresponding parameters, $t_*$ is a terminal moment. Integrating first equation of the system (\ref{eq:Phi}) we achieve:
\begin{equation}\label{eq:beta_int}
\omega = \omega_0 \exp\left[-\int\limits_{\beta_0}^{\beta_1}\frac{d\beta}{\beta-\Phi_1(\beta,\vartheta)}\right].
\end{equation}

It is important to mention, that function $\Phi_1(\beta,\vartheta)$ depends not only on $\beta$ and $\vartheta$ but also on the shape of the contact area, pressure distribution law $p(\xi,\eta)$, components of the friction matrix $f_x,\quad f_y$ and the angle $\varphi$ (orientation of the body on the surface). Thus, value of  $\beta_1$, when integral in (\ref{eq:beta_int}) becomes improper and seeks $-\infty$, depends on parameters of the mechanical system: 
\begin{equation}
\beta_1 \underset{\substack{t\to t_*\\ \vartheta\to\vartheta_*\\ \varphi\to\varphi_*}}\longrightarrow \beta_*(\vartheta_*, \varphi_*, \Omega, f_x, f_y, p(\xi,\eta)).
\end{equation}

Summarizing, note that by the time $t_*$  relation (\ref{eq:normal_force_rule})   and  
\begin{equation}\label{eq:Phi_beta}
\beta - \Phi_1(\beta,\vartheta) \underset{\substack{t\to t_*\\ \vartheta\to\vartheta_*\\ \beta \to\beta_*}}\longrightarrow 0.
\end{equation}
should be achieved.

Furthermore, with fixed values of $\beta = \tilde{\beta}$ equations $T_n(\tilde{\beta},\vartheta) = 0$ and (\ref{eq:Phi_beta}) may have several solutions. However, both conditions (\ref{eq:normal_force_rule}) and (\ref{eq:Phi_beta}) are achieved with singular $\vartheta_*,\quad \beta_*$  \cite{Dmitriev2002}, which depend on initial conditions.

It is important to mention that from system (\ref{eq:move_frenet_dimless}) with introducing parameter $\delta = \beta^{-1}$ we may derive the following system of equations:
\begin{equation}\label{eq:Phi2}
\begin{array}{l l}
\displaystyle
\frac{d\omega}{dv_C} = \Phi_2(\delta,\vartheta),\\
\displaystyle
v_C\frac{d\vartheta}{dt} = T_n(\delta,\vartheta).
\end{array}
\end{equation}
From system (\ref{eq:Phi2}) we achieve:
\begin{equation}\label{eq:normal_force_rule2}
T_n(\delta,\vartheta)\underset{\substack{t \to t_*\\ \vartheta\to\vartheta_*\\ \delta\to\delta_*}}\longrightarrow  0,\\ 
\end{equation}
and
\begin{equation}\label{eq:delta_int}
v_C = v_{C0} \exp\left[-\int\limits_{\delta_0}^{\delta_1}\frac{d\delta}{\Phi_2(\delta,\vartheta)-\delta}\right],
\end{equation} 
and, finally, with same reasoning:
\begin{equation}\label{eq:Phi2_delta}
\Phi_2(\delta,\vartheta)-\delta \underset{\substack{t\to t_*\\ \vartheta\to\vartheta_*\\ \delta \to\delta_*}}\longrightarrow 0.
\end{equation} 

During searching limit values of $\vartheta_*, \quad \beta_*$ with equations (\ref{eq:normal_force_rule}) and (\ref{eq:Phi_beta}), it may occur that there are no roots. Thus,  we should solve equations (\ref{eq:normal_force_rule2}) and (\ref{eq:Phi2_delta}) to find $\vartheta_*, \quad \delta_*$. Furthermore, because there is a strict dependence of plate motion
on interrelations between inertia moment and friction coefficients (see, for example, \cite{Dmitriev2009_4}) we should all the time check solution in both regions.

\subsection{Symmetric orthotropic friction. Specific initial conditions of motion}\label{subsec:res_sym}
\subsubsection{Initial conditions of motion $\omega=0, \quad v \neq 0$}\label{subsubsec:res_omegaeq0}

Let's get back to system (\ref{eq:frenet}). With (\ref{eq:velocity}) and (\ref{eq:forces_xy}) it is possible to show that in case $\omega=0, \quad v\neq0$ we have:
\begin{equation}\label{eq:case_sym_omega0_elemfric}\nonumber
\begin{array}{l l}
\displaystyle
\tau_x = -p \left(f_x \cos\vartheta + f\sin\vartheta\right),\\
\tau_y = -p\left(-f\cos\vartheta + f_y\sin\vartheta\right),
\end{array}
\end{equation}
and, thus, we achieve system:
\begin{equation}\label{eq:case_sym_omega0_move}
\begin{array}{l l}
\displaystyle
m\dot v_C = T_x\cos\vartheta + T_y\sin\vartheta = -p\iint\limits_{\Omega}\left(\mu\sin^2\vartheta + f_x\right)d\xi d\eta,\\
\displaystyle
mv_C\dot \vartheta = -T_x\sin\vartheta + T_y\cos\vartheta = -p\iint\limits_{\Omega}\left(\mu\sin\vartheta\cos\vartheta - f\right)d\xi d\eta,\\
\displaystyle
I\dot\omega = \iint\limits_{\Omega}\left(\tau_y x' - \tau_x y'\right)d\xi d\eta = -p\iint\limits_{\Omega}\left(\xi K_1+\eta K_2\right)d\xi d\eta,
\end{array}
\end{equation}
where
$$K_1 = (\mu\sin\vartheta\cos\varphi + f_x\sin(\vartheta-\varphi)-f\cos(\vartheta-\varphi)),$$
$$K_2 = (-\mu\sin\vartheta\sin\varphi - f_x\cos(\vartheta-\varphi) - f\cos(\vartheta-\varphi)).$$
It can be shown that integrals in the third equation of system (\ref{eq:case_sym_omega0_move}) is equal to zero for elliptical and circular contact areas.
So, finally we have:
\begin{equation}\label{eq:case_sym_omega0_finmove}
\begin{array}{l l}
\displaystyle
m\dot v_C = -p S \left(\mu\sin^2\vartheta + f_x\right),\\
m v_C\dot \vartheta = -p S \left(\mu\sin\vartheta\cos\vartheta - f\right),\\
I\dot\omega = 0.
\end{array}
\end{equation}

From the  system of equations (\ref{eq:case_sym_omega0_finmove}) we see, that in case the initial motion is translational it stays translational until the final point.

\subsubsection{Initial conditions of motion $\omega \neq 0, \quad v = 0$}\label{subsec:res_veq0}

For equation  (\ref{eq:frenet}) taking into account conditions $v=0, \quad \omega\neq0$ we have that parameter $\beta$ is zero and from (\ref{eq:velocity}) and (\ref{eq:forces_xy}) we achieve system of equations:
\begin{equation}\label{eq:case_sym_v0_move}
\begin{array}{l l}
\displaystyle
m\dot v_C = -p \iint\limits_{\Omega} \frac{\xi A_0 + \eta A_1}{D}d\xi d\eta = -p \left[A_0 F_1(\xi,\eta) + A_1 G_1(\xi,\eta)\right],\\
\displaystyle
m v_C\dot\vartheta = -p \iint\limits_{\Omega} \frac{\xi B_0 + \eta B_1}{D}d\xi d\eta  = -p \left[ B_0 F_1(\xi,\eta) + B_1 G_1(\xi,\eta)\right],\\
\displaystyle
I\dot\omega = -p\iint\limits_{\Omega}\frac{\xi^2 C_0 + \xi\eta C_1 + \eta^2 C_2}{D}d\xi d\eta = -p\left[C_0 F_2(\xi,\eta) + C_1 L(\xi,\eta) + C_2 G_2(\xi,\eta)\right],
\end{array}
\end{equation}
where
$$D = \sqrt{\xi^2+\eta^2},$$
$$A_0 = \mu\cos\varphi\sin\vartheta + f_x\sin(\vartheta-\varphi) + f\cos(\vartheta-\varphi),$$
$$A_1 = f\sin(\vartheta-\varphi) - \mu\sin\vartheta\sin\varphi - f_x\cos(\vartheta-\varphi),$$
$$B_0 = \mu\cos\varphi\cos\vartheta + f_x\cos(\vartheta-\varphi) - f\sin(\vartheta-\varphi),$$
$$B_1 = f\cos(\vartheta - \varphi) + f_x\sin(\vartheta-\varphi)-\mu\sin\varphi\cos\vartheta,$$
$$C_0 = f_x+\mu\cos^2\varphi,\quad C_1 = -2\mu\cos\varphi\sin\varphi,\quad C_2 = f_x+\mu\sin^2\varphi.$$
$$F_1(\xi,\eta)=\iint\limits_{\Omega} \frac{\xi}{\sqrt{\xi^2+\eta^2}}d\xi d\eta,\quad G_1(\xi,\eta)= \iint\limits_{\Omega} \frac{\eta}{\sqrt{\xi^2+\eta^2}}d\xi d\eta ,$$
$$F_2(\xi,\eta)=\iint\limits_{\Omega} \frac{\xi^2}{\sqrt{\xi^2+\eta^2}}d\xi d\eta ,\quad G_2(\xi,\eta)= \iint\limits_{\Omega} \frac{\eta^2}{\sqrt{\xi^2+\eta^2}}d\xi d\eta ,$$
$$L(\xi,\eta)= \iint\limits_{\Omega} \frac{\xi\eta}{\sqrt{\xi^2+\eta^2}}d\xi d\eta .$$

Now we find integrals $F_1(\xi,\eta)$, $G_1(\xi,\eta)$, $F_2(\xi,\eta)$, $G_2(\xi,\eta)$, $L(\xi,\eta)$. 
It can be seen that for  $F_1(\xi,\eta)$, $G_1(\xi,\eta)$ and $L(\xi,\eta)$ the integration area is symmetrical, thus:
\begin{equation}\label{eq:case_sym_v0_FG_1}
\begin{array}{l l l}
\displaystyle
F_1(\xi,\eta) = \int\limits_{-b}^{b}d\eta\int\limits_{-a h_2(\eta)}^{a h_2(\eta)}\frac{\xi}{\sqrt{\xi^2+\eta^2}}d\xi = 0,\quad
G_1(\xi,\eta) = \int\limits_{-a}^{a}d\xi\int\limits_{-b h_1(\xi)}^{b h_1(\xi)}\frac{\eta}{\sqrt{\xi^2+\eta^2}}d\eta = 0,\\
\displaystyle
L(\xi,\eta) = \int\limits_{-b}^{b}\eta d\eta\int\limits_{-a h_2(\eta)}^{a h_2(\eta)}\frac{\xi}{\sqrt{\xi^2+\eta^2}}d\xi = 0,\quad \displaystyle h_1 = \sqrt{1-\frac{\xi^2}{a^2}}, \quad \displaystyle h_2 = \sqrt{1 - \frac{\eta^2}{b^2}}.\\ 
\end{array}
\end{equation}
Integrals  $F_2(\xi,\eta)$, $G_2(\xi,\eta)$ are not equal to zero for the area. 

\begin{equation}\label{eq:case_sym_v0_FG_2}
\begin{array}{l l}
\displaystyle
F_2(\xi,\eta) = \int\limits_{-b}^{b}d\eta\int\limits_{-a h_2(\eta)}^{a h_2(\eta)} \frac{\xi^2}{\sqrt{\xi^2+\eta^2}}d\xi =  -\int\limits_{-b}^{b}\frac{\eta^2}{2}\ln\frac{\left(\sqrt{a^2h_2^2(\eta)+\eta^2}+ ah_2(\eta)\right)^2}{\eta^2}d\eta,\\
\displaystyle
G_2(\xi,\eta) = \int\limits_{-a}^{a}d\xi\int\limits_{-b h_1(\xi)}^{b h_1(\xi)}\frac{\eta^2}{\sqrt{\xi^2+\eta^2}}d\eta =  -\int\limits_{-a}^{a}\frac{\xi^2}{2}\ln\frac{\left(\sqrt{b^2h_1^2(\xi)+\xi^2}+ bh_1(\xi)\right)^2}{\xi^2}d\xi.
\end{array}
\end{equation}

Thus equations(\ref{eq:case_sym_v0_move}) transform into:
\begin{equation}\label{eq:case_sym_v0_move_fin}
\begin{array}{l l l}
\displaystyle
m\dot v_C = 0,\\
\displaystyle
m v_C\dot\vartheta = 0,\\
\displaystyle
I\dot\omega = C_0 F_2(\xi,\eta) + C_2 G_2(\xi,\eta).
\end{array}
\end{equation}

For symmetric orthotropic friction and uniform pressure distribution it can be stated that if the initial motion is rotational it stays rotational until the end.


\subsection{Symmetric and asymmetric orthotropic friction. Numerical results}\label{subsec:res_asym}

If the initial values of angular and linear (sliding) velocities are non-zero any analytic simplifications can hardly be reached. In our paper \cite{Silantyeva2014} system of equations (\ref{eq:move_frenet_dimless}) was solved in ($\xi, \eta$) coordinate system, but with this approach an accuracy of results was not enough satisfied. Furthermore, at the most final points the method tends to give  significant oscillations, because near $\beta = \beta_*$ situation with singularity may appear. So we switched to the method Lurye described here. After manipulations with forces integrals according to the described method, we obtained a very stable and accurate numerical procedure.  With this new approach we achieved  numerical solution for uniform pressure distribution and symmetric and asymmetric friction forces.

Table \ref{tab:asym_beta_theta_area} shows resultant values of parameters of interest for circular and elliptic plates.  The instantaneous velocity center for both types of plates is located in the same area. However, velocity vector rotational angle $\vartheta_*$ is noticeably higher for elliptic plate. Instantaneous velocity center position described by $\beta_*$ is lower for elliptic plate comparing with circular one. 

In our paper \cite{Silantyeva2016_vestnik} results for symmetric orthotropic case were presented. For symmetric case $\beta_*$ is  significantly lower for both circular and elliptic plates comparing to asymmetric example. Furthermore, for circular plate in symmetric orthotropic friction  we have $\vartheta_* = 0$. However, in asymmetric case for both shapes $\vartheta_*$ values show that velocity vector orients to the 3rd quadrant, which is actually the one with the lowest coefficients of friction.

\begin{table}[!th] 
\centering
\caption{Parameters $\beta_*, \quad \vartheta_*$ for circular and elliptical plates for asymmeric orthotropic friction ($\vartheta_0=\frac{\pi}{4},\quad \varphi_0=\frac{\pi}{3}$)}\label{tab:asym_beta_theta_area}
\begin{tabular}{|p{0.7cm}|p{1.2cm}|p{1.2cm}|p{1.2cm}|p{1.2cm}|p{1.2cm}|p{1.2cm}|}
\hline\noalign{\smallskip}
&\multicolumn{3}{|c|}{Circle} & \multicolumn{3}{|c|}{Ellipse ($e=0.6$)}  \\
\hline\noalign{\smallskip}
$\mu_+$& $\beta_*$& $\vartheta_*$& Area&$\beta_*$ & $\vartheta_*$&Area\\ 
\hline\noalign{\smallskip}
0.03& 0.887 & -2.46&13 &0.81&-2.71& 13\\
\hline\noalign{\smallskip}
0.06& 0.908 & -2.57&13 &0.83&-2.77& 13\\
\hline\noalign{\smallskip}
0.09& 0.937 & -2.65&13 &0.86&-2.82& 13\\
\hline\noalign{\smallskip}
0.12& 0.976 & -2.71&13 &0.89&-2.86& 13\\
\hline\noalign{\smallskip}
0.15& 1.042 & -2.78&12 &0.91&-2.88& 12\\ 
\hline\noalign{\smallskip}
0.18& 1.197 & -2.86&19 &0.99&-2.93& 19\\
\hline\noalign{\smallskip}
\end{tabular}
\end{table}

\begin{figure}[!th]
\includegraphics[width=14cm]{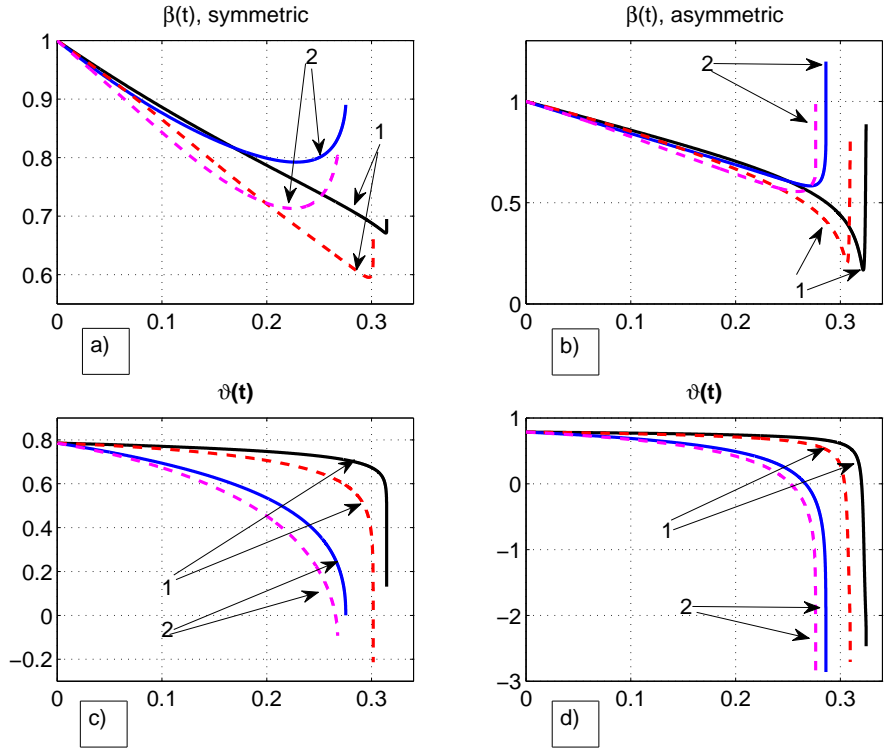}
\caption{Parameters $\beta$ and $\vartheta$ evolution for circular (solid line) and elliptic (dashed line) plate for orthotropic friction: (1) $\mu = \mu_+ = 0.03$, (2) $\mu = \mu_+ = 0.18$}\label{fig:8:asym_sym_compare}
\end{figure}

Figure \ref{fig:8:asym_sym_compare} demonstrates evolution of $\beta(t), \quad \vartheta(t)$ for symmetric and asymmetric cases. It was assumed that  $a=1$ and represents major semi-axes of  elliptic plate, which eccentricity is $e = 0.6$ and for circular plate $e=0$. Initial conditions taken in the example are: $v_0 = 1, \quad \omega_0 = 1,\quad \vartheta_0 = \pi/4$, and initial ellipse orientation angle is $\varphi_0 = \pi/3$. Friction coefficient   $\mu_+ = f_{y+} - f_{x+}$ with $f_x = f_{x+} = 0.42, \quad f_{x-} = 0.5f_{x+}, f_y = f_{y+},\quad f_{y-} = 0.5 f_{y+}$.  For both symmetric and asymmetric cases sliding and spinning end simultaneously. This important outcome was also achieved for non-uniform pressure distributions: for circular plate with respect to isotropic friction force and   axisymmetric normal pressure  in \cite{Argatov2005_en}, under linear pressure distribution  in \cite{Borisov2015_en}, and for elliptic plate under linear pressure distribution and symmetric orthotropic friction in \cite{Silantyeva2014}. However, figure shows significant difference in behavior of $\beta(t),\quad \vartheta(t)$ curves with respect to asymmetry of friction force. Shape factor is noticeably important for both symmetric and asymmetric cases: elliptic plate moves shorter period of time with more velocity vector rotation  (higher changes of $\vartheta$ values). 

\section{Conclusion}\label{sec:conclusion}
\begin{itemize}
\item Problem of terminal motion of thin elliptic plate taking into account anisotropy of friction force is formulated. General analytic results show that it is possible to analyze  the system of motion equations without specifying pressure distribution. It is stated that until the terminal point two conditions (\ref{eq:Phi_beta}) and (\ref{eq:normal_force_rule}) should be achieved  simultaneously. 
\item Total friction force and total moment evaluations are shown in the case of thin elliptic plate  under uniform pressure distribution. Two cases are discussed: symmetric orthotropic friction and asymmetric orthotropic friction. Based on the Lurye method friction force is obtained for both cases. This leads to significant simplifications of numerical procedure.
\item Some specific cases of initial conditions are analyzed separatelywith the aid of numerical study.   It is shown, that for symmetric orthotropic friction and elliptic contact area  under uniform pressure distribution: if the initial motion  is linear it stays linear till the end. In case initial motion is rotational it is rotational during the whole period of motion. 
\item Numerical results are presented for symmetric and asymmetric friction forces and elliptic and circular contact areas under uniform pressure distribution. It is stated that sliding and spinning end simultaneously both in symmetric and asymmetric cases. Figures show significant influence of asymmetry of friction forces on the motion. Contact area impact is also pointed out.
\end{itemize}

\bibliographystyle{acm}
\bibliography{silantyeva_zamm2017_bib}

\providecommand{\WileyBibTextsc}{}
\let\textsc\WileyBibTextsc
\providecommand{\othercit}{}
\providecommand{\jr}[1]{#1}
\providecommand{\etal}{~et~al.}


\begin{thebibliography}{[10]}

\bibitem{Antoni2007}
 \textsc{N.~Antoni},  \textsc{J.\,L. Ligier},  \textsc{P.~Saffre},  and
  \textsc{J.~Pastor},
Asymmetric friction: Modeling and experiments.,
 \jr{Int.J.Eng. Sci.} \textbf{45}, 587--600 (2007).


\bibitem{Argatov2005_en}
 \textsc{I.~Argatov},
Equilibrium conditions for a rigid body on a rough plane in the case of axially
  symmetric distribution of normal pressures,
 \jr{Mech. Solids} \textbf{40}(2), 11--20 (2005).


\bibitem{Bafekrpour2015}
 \textsc{E.~Bafekrpour},  \textsc{A.~Dyskin},  \textsc{E.~Pasternak},
  \textsc{A.~Molotnikov},  and  \textsc{Y.~Estrin},
Internally architectured materials with directionally asymmetric friction,
 \jr{Scientific Reports} \textbf{5}, 10732 EP (2015).


\bibitem{Borisov2015_en}
 \textsc{A.~Borisov},  \textsc{Y.~Karavaev},  \textsc{I.~Mamaev},
  \textsc{N.~Erdakova},  \textsc{T.~Ivanova},  and  \textsc{V.~Tarasov},
Experimental investigation of the motion of a body with an axisymmetric base
  sliding on a rough plane,
 \jr{Regular and Chaotic Dynamics} \textbf{20}(5), 518--541 (2015).


\bibitem{Campione2012}
 \textsc{M.~Campione},  \textsc{S.~Trabattoni},  and  \textsc{M.~Moret},
Nanoscale mapping of frictional anisotropy,
 \jr{Tribol Lett} \textbf{45}(219-224) (2012).


\othercit
\bibitem{Carbone2009}
 \textsc{G.~Carbone},  \textsc{A.~Malchikov},  \textsc{M.~Ceccarelli},  and
  \textsc{S.~Jatsun},
Design and simulation of kursk robot for in-pipe inspection,
 in: Proceedings of the 10th IFToMM International Symposium on Science of
  Mechanisms and Machines,  (Springer, Brasov, Romania, October 12-15 2009),
  pp.\,103--114.


\bibitem{Dmitriev2002}
 \textsc{N.~Dmitriev},
Movement of the disk and the ring over the plane with anisotropic friction.,
 \jr{J.Fric. Wear} \textbf{23}, 10--15 (2002).


\bibitem{Dmitriev2009_4}
 \textsc{N.~Dmitriev},
Sliding of a solid body supported by a round platform on a horizontal plane
  with orthotropic friction. part~1. regular load distribution.,
 \jr{J.Fric. Wear} \textbf{30}(4), 227--234 (2009).


\bibitem{Dmitriev2013}
 \textsc{N.~Dmitriev},
Motion of material point and equilibrium of two-mass system under asymmetric
  orthotropic friction.,
 \jr{J.Fric. Wear} \textbf{34}, 429--437 (2013).


\bibitem{Dmitriev2015}
 \textsc{N.~Dmitriev},
Motion of a narrow ring on a plane with asymmetric orthotropic friction.,
 \jr{J.Fric. Wear} \textbf{36}, 80--88 (2015).


\othercit
\bibitem{Silantyeva2014}
 \textsc{N.~Dmitriev} and  \textsc{O.~Silantyeva},
About the movement of a solid body on a plane surface in accordance with
  elliptic contact area and anisotropic friction force,
 in: Proc. of jointly organised WCCM XI, ECCM V, ECFD VI. Vol. IV,  (CIMNE,
  Barcelona, Spain, 2014),  pp.\,4440--4452.


\bibitem{Silantyeva2016_vestnik}
 \textsc{N.~Dmitriev} and  \textsc{O.~Silantyeva},
Terminal motion of a thin elliptical plate over a horizontal plane with
  orthotropic friction,
 \jr{Vestnik St. Petersburg University: Mathematics} \textbf{49}(1), 92--98
  (2016).


\bibitem{Farkas2003}
 \textsc{Z.~Farkas},  \textsc{G.~Bartels},  \textsc{T.~Unger},  and
  \textsc{D.~Wolf},
Frictional coupling between sliding and spinning motion,
 \jr{Phys. Rev. Lett.} \textbf{90}(24), 248--302 (2003).


\bibitem{Ishigami2012}
 \textsc{G.~Ishigami},  \textsc{J.~Overholt},  and  \textsc{K.~Iagnemma},
Multi-material anisotropic friction wheels for omnidirectional ground vehicles,
 \jr{Journal of Robotics and Mechatronics} \textbf{24}, 261--267 (2012).


\bibitem{Konyukhov2008}
 \textsc{A.~Konyukhov},  \textsc{P.~Vielsack},  and  \textsc{K.~Schweizerhof},
On coupled models of anisotropic contact surfaces and their experimental
  validation.,
 \jr{Wear} \textbf{264}, 579--588 (2008).


\bibitem{Lopes2015}
 \textsc{D.\,S. Lopes},  \textsc{R.\,R. Neptune},  \textsc{J.\,A.
  Ambr{\'o}sio},  and  \textsc{M.\,T. Silva},
A superellipsoid-plane model for simulating foot-ground contact during human
  gait,
 \jr{Computer Methods in Biomechanics and Biomedical Engineering} pp.\,1--10
  (2015).


\othercit
\bibitem{Lurye2002}
 \textsc{A.~Lurye},
Analytical Mechanics (Springer-Verlag Berlin Heidelberg, Berlin, 2002).


\bibitem{Piotrowski2005}
 \textsc{J.~Piotrowski} and  \textsc{H.~Chollet},
Wheel-rail contact models for vehicle system dynamics including multi-point
  contact,
 \jr{Vehicle Syst. Dyn.} \textbf{43}(6-7), 455--483 (2005).


\othercit
\bibitem{Rozenblat2006}
 \textsc{G.~Rozenblat},
Dynamical Systems with Dry Friction. (NITs Regular and Chaotic Dynamics,
  Moscow-Izhevsk, 2006).


\othercit
\bibitem{Silantyeva2016_gdansk}
 \textsc{O.~Silantyeva} and  \textsc{N.~Dmitriev},
Dynamics of bodies under symmetric and asymmetric orthotropic friction forces,
 in: Advances in Mechanics: Theoretical, Computational and Interdisciplinary
  Issues,  (CRC Press/Balkema, Taylor \& Francis Group, eds. M. Kleiber et al.,
  London, 2016), chap.~106,  pp.\,511--515.


\bibitem{Weidman2007}
 \textsc{P.~Weidman} and  \textsc{C.~Malhotra},
On the terminal motion of sliding spinning disks with uniform coulomb
  friction.,
 \jr{Phys. D} \textbf{233}(1), 1--13 (2007).


\bibitem{Zmitrowicz1989}
 \textsc{A.~Zmitrowicz},
Mathematical descriptions of anisotropic friction.,
 \jr{Int.J.Solids Struct.} \textbf{25}(8), 837--862 (1989).


\bibitem{Zmitrowicz1992_ex}
 \textsc{A.~Zmitrowicz},
￼illustrative examples of centrosymmetric and non-centrosymmetric anisotropic
  friction.,
 \jr{Int.J.Solids Struct.} \textbf{29}(23), 3045--3059 (1992).


\bibitem{Zmitrowicz2005}
 \textsc{A.~Zmitrowicz},
Models of kinematic dependent anisotropic and heterogeneous friction.,
 \jr{Int.J.Solids Struct.} \textbf{43}, 4407--4451 (2005).


\bibitem{Zmitrowicz2010}
 \textsc{A.~Zmitrowicz},
Contact stresses: a short survey of models and methods of computations,
 \jr{Arch App Mech} \textbf{80}, 1407--1428 (2010).


\end{thebibliography}

\end{document}